\documentclass[11pt]{article}

\usepackage[preprint]{acl}

\usepackage{times}
\usepackage{latexsym}

\usepackage[T1]{fontenc}

\usepackage[utf8]{inputenc}

\usepackage{microtype}

\usepackage{inconsolata}

\usepackage{graphicx}

\usepackage{multirow}
\usepackage{longtable}
\usepackage{xspace}
\usepackage{enumitem}
\usepackage{multicol}
\usepackage{amssymb}
\usepackage{amsmath}
\usepackage{graphicx}
\usepackage{xcolor}
\usepackage{colortbl}
\usepackage{amsfonts} 
\usepackage{multirow}
\usepackage{natbib}
\usepackage{url}
\usepackage{algorithm}
\usepackage[noend]{algpseudocode}

\usepackage{algorithmicx}

\usepackage{caption}
\usepackage{comment}
\usepackage{cleveref}
\usepackage{subcaption}
\usepackage[table]{xcolor}
\usepackage{colortbl}

\algrenewcommand\algorithmicindent{0.8em}
\algrenewcommand\alglinenumber[1]{\scriptsize #1}

\newcommand{\IllusionAudio}{\textsc{AI-CAPTCHA}\xspace}
\newcommand{\IllusionAudioEval}{\textsc{ACEval}\xspace}
\newcommand{\IllusionAudioScheme}{\textsc{IllusionAudio}\xspace}
\newcommand{\Captcha}{\textsc{CAP\-TCHA}\xspace}

\definecolor{lightgreen}{RGB}{210,235,220}

\title{Robust CAPTCHA Using Audio Illusions in the Era of Large Language Models: from Evaluation to Advances}

\author{
\begin{tabular}{c}
Ziqi Ding \quad
Yunfeng Wan \quad
Wei Song \quad
Yi Liu \quad
Gelei Deng \\[0.3em]
Nan Sun \quad
Huadong Mo \quad
Jingling Xue \quad
Shidong Pan \quad
Yuekang Li
\end{tabular}
}

\begin{document}
\maketitle
\begin{abstract}
CAPTCHAs are widely used by websites to block bots and spam by presenting challenges that are easy for humans but difficult for automated programs to solve.
To improve accessibility, audio CAPTCHAs are designed to compliment visual ones.
However the robustness of audio CAPTCHAs against advanced Large Audio Language Models (LALMs) and Automatic Speech Recognition (ASR) models remains unclear.
In this paper, we introduce \IllusionAudio, a unified framework that offers (i) an evaluation framework, \IllusionAudioEval, which includes advanced LALM- and ASR-based solvers, and (ii) a novel audio CAPTCHA approach, \IllusionAudioScheme, leveraging audio illusions.
Through extensive evaluations of seven widely deployed CAPTCHAs, we show that most methods can be solved with high success rates by the advanced LALMs and ASR models, exposing critical security weaknesses.
To address this, we design a new CAPTCHA approach, \IllusionAudioScheme, which exploits perceptual illusion cues rooted in human auditory mechanisms.
Extensive experiments demonstrate that our method defeats all tested LALMs- and ASR-based attacks while achieving a 100\% human pass rate, significantly outperforming existing methods.

\end{abstract}

\section{Introduction}
\label{sec:introduction}

Completely Automated Public Turing Tests to Tell Computers and Humans Apart (\Captcha{}s) are widely used to protect online services from automated bot attacks~\citep{von2003captcha,gossweiler2009s}.
Most deployed \Captcha{}s rely on vision-based challenges~\citep{Ding2025NGCaptchaAC,ding2025illusioncaptcha}, such as distorted text or image recognition. 
While effective, these visual challenges are inaccessible to people with visual impairments (PVIs)~\citep{fanelle2020blind,shirali2007captcha}.
To address this accessibility gap, audio \Captcha{}s were introduced as an alternative modality, enabling users to solve auditory challenges instead~\citep{gao2010audio,alnfiai2020novel}.

Existing audio \Captcha{}s~\citep{szegedy2017inception,he2016deep,hunt2014artificial,fanelle2020blind,arkoselabsArkoseMatchKey} can be broadly categorized into two types: content-based and rule-based.
Content-based audio \Captcha{}s~\citep{abdullah2022attacks,aubry2025bypassing,tam2008breaking} require users to listen to spoken words or digits and transcribe the content~\citep{fanelle2020blind}. These schemes assume that speech perception and transcription are straightforward for humans but challenging for automated systems.
To enhance robustness, many deployed content-based audio \Captcha{}s incorporate acoustic perturbations, such as background noise, distortions, or overlapping sounds.
In contrast, rule-based audio \Captcha{}s~\citep{fanelle2020blind,arkoselabsArkoseMatchKey} move beyond transcription, requiring users to follow audio instructions or identify specific sound events.
This design tests higher-level auditory reasoning rather than raw speech recognition.

Existing audio \Captcha{}s effectively counter traditional bots~\citep{fanelle2020blind}. 
However, it is unclear whether their security assumptions remain valid given recent advances in Automatic Speech Recognition (ASR) and Large Audio Language Models (LALMs). 
Modern ASR systems achieve near-human performance on noisy and distorted speech, and LALMs further extend these capabilities by reasoning over complex audio inputs. 
These advances weaken the assumption that audio \Captcha{}s can reliably distinguish humans from automated solvers. 
Tasks such as transcribing corrupted speech and interpreting audio instructions may no longer separate human users from AI-driven systems.
Therefore, systematic evaluation of current audio \Captcha{}s against advanced models is necessary. 
If performance is inadequate, new audio \Captcha{} designs should be developed.

To address the research gaps, we introduce \IllusionAudio, a unified framework comprising two components: (1)  \IllusionAudioEval, an evaluation framework for systematically assessing the robustness of audio \Captcha{} schemes against advanced LALM-based and ASR-based solvers, and (2) \IllusionAudioScheme, a novel audio \Captcha{} design that leverages audio illusions.

\IllusionAudioEval incorporates two types of AI-driven solvers.
The first is an LALM-based solver that directly reasons over audio input to generate answers, representing a new class of attacks against audio \Captcha{}s.
The second is an ASR-based solver that follows a two-stage pipeline: audio is first transcribed by an ASR model, and the resulting transcript is then processed by a Large Language Model (LLM) to produce the final response. Both solvers ultimately produce an audio-based answer, rather than relying solely on ASR to transcribe the audio.

Using \IllusionAudioEval, we assess seven deployed audio \Captcha{} schemes, including four content-based (Geetest, Google, MTCaptcha, and Telephone audio \Captcha{}s) and three rule-based (Math, Character, and Arkose Labs audio \Captcha{}s).
We assess their robustness against three advanced LALMs (Qwen-Audio-Chat, SeaLLMs-Audio-7B, and Qwen2-Audio-7B-Instruct) under zero-shot and chain-of-thought prompting, and two leading ASR models (GPT-4o-Transcript and GPT-4o-mini-Transcript) under prompt-guided and non-prompt-guided settings, with GPT-4o used for downstream reasoning.
In addition, we conduct a user study with 63 human participants to assess the usability of these audio \Captcha{}s by measuring human success rates across multiple attempts.

Our evaluation yields two key findings. First, both LALM-based and ASR-based solvers achieve high bypass rates (41.67\% and 49.99\%, respectively) against existing deployed audio \Captcha{}s, highlighting their vulnerabilities to AI-driven attacks.
Second, most existing audio \Captcha{} schemes impose substantial difficulty on human users, with an average first-attempt success rate of only 61.90\%.

Based on these findings, we introduce \IllusionAudioScheme, a new audio \Captcha{} scheme that leverages the sine-wave speech illusion---a perceptual phenomenon in which humans can recognize speech from sparse acoustic cues that lack explicit linguistic structure.
While such signals remain intelligible to humans, they are challenging for LALM-based and ASR-based solvers to interpret. Beyond the illusion itself, \IllusionAudioScheme incorporates additional mechanisms, such as irreversible audio transformations, to further resist automated solvers.

Extensive experiments using our evaluation framework demonstrate that \IllusionAudioScheme{} defeats all tested LALM-based and ASR-based solvers, achieving a 0\% bypass rate.
A user study with 63 participants, including PVIs, shows that \IllusionAudioScheme achieves a 100\% first-attempt success rate, substantially outperforming existing audio \Captcha{} schemes in both security and usability.

In summary, we make the following contributions:
\begin{itemize}[leftmargin=*, topsep=0pt, itemsep=0pt]

\item We introduce \IllusionAudio, a unified framework comprising \IllusionAudioEval, an evaluation framework that employs advanced LALM-based and ASR-based solvers to assess audio \Captcha{} robustness, and \IllusionAudioScheme, a novel audio \Captcha{} design based on audio illusions.

\item Using \IllusionAudioEval, we conduct a comprehensive evaluation of seven widely deployed audio \Captcha{} schemes, revealing critical security vulnerabilities and usability limitations in existing designs.

\item We propose \IllusionAudioScheme, which leverages sine-wave speech illusions to create a perceptual gap between humans and AI. Experimental results demonstrate that \IllusionAudioScheme achieves 0\% bypass rate against all tested AI solvers while maintaining 100\% first-attempt human success rate.

\end{itemize}

\noindent \textbf{Ethical Considerations.} Our research focuses on the security and usability of audio \Captcha{}s. All user studies were approved by our institutional review board and raise no ethical concerns. We develop \IllusionAudio to enhance web security by effectively distinguishing human users from bots. Details of our research and ethical declaration are provided on our website~\citep{googleAICAPTCHA}: \url{https://sites.google.com/view/aicaptcha/}.

\section{Background and Related Work}
\label{sec:background}

\subsection{Audio \Captcha{}s} 

\begin{figure}[!t]
	\centering
    \includegraphics[width=\linewidth]{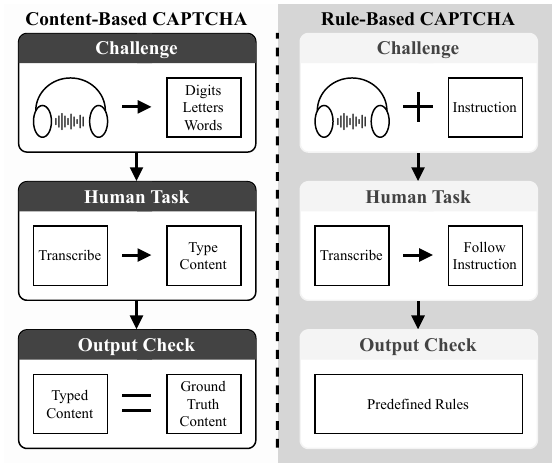}
	\caption{Architecture of content-based Audio \Captcha{}s and rule-based Audio \Captcha{}s.}
	\label{fig:background}
    \vspace*{-2ex}
\end{figure}

Audio \Captcha{}s are a class of challenge-response tests that leverage human auditory perception to defend against automated abuse~\citep{gao2010audio,fanelle2020blind,alnfiai2020novel}.
They were introduced as an accessible alternative to visual \Captcha{}s and have become an important component of usable web security~\citep{saini2013review}. As illustrated in \Cref{fig:background}, existing audio \Captcha{} schemes can be categorized into two types: content-based and rule-based. Content-based audio \Captcha{}s~\citep{abdullah2022attacks,aubry2025bypassing,tam2008breaking} typically require users to transcribe spoken digits or letters.
In contrast, rule-based audio \Captcha{}s~\citep{fanelle2020blind,arkoselabsArkoseMatchKey} require users to reason over both audio content and textual instructions, shifting the challenge from simple transcription toward higher-level audio reasoning.
While these designs improve resistance to traditional automated attacks, their effectiveness against advanced AI-driven solvers (e.g., LALMs) has not been systematically examined.

\subsection{Large Audio Language Models}

Recent advances in LLMs have provided empirical evidence for scaling laws, reshaping how intelligent systems are trained and evaluated~\citep{achiam2023gpt,team2023gemini,gu2024survey,ge2023openagi}.
By leveraging large-scale data and advanced training techniques~\citep{liu2021makes}, modern LLMs exhibit strong capabilities in perception and reasoning.
Building on this foundation, LALMs extend LLMs to the audio domain by jointly modeling speech and text, enabling reasoning over complex audio inputs and multimodal instructions~\citep{Qwen-Audio,Qwen2-Audio,SeaLLMs-Audio}.
Advanced LALMs, such as Qwen-Audio-Chat, SeaLLMs-Audio-7B, and Qwen2-Audio-7B-Instruct, have demonstrated strong performance on various audio understanding tasks~\citep{wang2025audio,wang2024audiobench,yang2024air}, including audio question answering and audio-text reasoning.

\subsection{Automatic Speech Recognition} 

ASR systems are designed to transcribe audio signals into text. Early approaches relied on modular pipelines combining Hidden Markov Models (HMMs) with Deep Neural Networks (DNNs)~\citep{Li2021RecentAI,Nayeem2025AutomaticSR}. Subsequent research shifted toward end-to-end frameworks, including Connectionist Temporal Classification (CTC), attention-based encoder--decoder architectures, and transducer-based approaches~\citep{Bahdanau2014NeuralMT,Chorowski2015AttentionBasedMF}. Modern ASR systems leverage deep learning and large-scale datasets to achieve accurate and robust speech-to-text transcription.
These systems can now handle a wide range of speech patterns and accents, making them effective at transcribing audio \Captcha{} challenges.

\subsection{Audio Illusions} 

Audio illusions~\citep{deutsch1974auditory,tiippana2014mcgurk,mitIllusionsTrick} are perceptual phenomena in which human listeners experience sounds in ways that diverge from their physical acoustic structure.
Such illusions arise from the human auditory system's ability to infer structure and meaning beyond explicit signal cues.
Representative examples include the \emph{McGurk effect}, where conflicting visual and auditory inputs alter perceived speech~\citep{tiippana2014mcgurk}, and the \emph{Shepard tone}, which creates the illusion of an endlessly rising pitch from a finite signal~\citep{shepherd2017musical}.
Particularly relevant to our work is the \emph{sine-wave speech illusion}, in which natural speech is reduced to sparse sinusoidal components. These signals remain intelligible to humans despite lacking the conventional spectral features that AI models typically rely on~\citep{mitIllusionsTrick}.
This perceptual asymmetry, where humans can interpret signals that AI systems cannot, enables the construction of audio \Captcha{}s that are solvable by humans yet challenging for automated solvers.

\subsection{Related Work}

\noindent \textbf{Audio \Captcha{} Attacks.}
Prior work has demonstrated that traditional audio \Captcha{}s are vulnerable to ASR-based attacks~\citep{tam2008breaking,sano2013solving}. These attacks typically transcribe distorted speech using noise-robust ASR models.
More recent work has shown that modern ASR systems can bypass many deployed audio \Captcha{}s despite noise injection~\citep{abdullah2022attacks,aubry2025bypassing}. 
However, these attacks only transcribe the audio rather than give the answer of audio \Captcha{}s.
Moreover, they do not hold the reasoning capabilities needed to solve the rule-based \Captcha{}s.
Therefore, the design of the solver in \IllusionAudioEval extends this line of research by evaluating the emerging threat of LALMs, which combine speech recognition with semantic reasoning, and add downstream reasoning function provided by LLMs for usual ASR systems.

\smallskip
\noindent \textbf{Perceptual \Captcha{} Designs.}
Researchers have explored perceptual phenomena to create \Captcha{}s that exploit human cognitive advantages.
In the visual domain, IllusionCAPTCHA~\citep{ding2025illusioncaptcha} leverages visual illusions to create challenges that are easy for humans but difficult for vision models.
Our work extends this concept to the audio domain, using sine-wave speech illusions to create a similar perceptual gap.
Unlike adversarial perturbations that add imperceptible noise to confuse models~\citep{zhang2021generating}, our approach fundamentally transforms the signal in a way that preserves human intelligibility while removing features that AI models rely on.

\smallskip
\noindent \textbf{Accessibility in \Captcha{}s.}
Audio \Captcha{}s were originally designed for accessibility, particularly for PVIs~\citep{fanelle2020blind,shirali2007captcha}.
However, studies have shown that many audio \Captcha{}s impose significant usability burdens on all users~\citep{shi2020text,kulkarni2018audio}.
Our work aims to address both security and accessibility by designing an audio \Captcha{} that is both robust against AI attacks and easy for humans including PVIs to solve.

Given these advances in AI capabilities and the limitations of existing \Captcha{} designs, we next formalize the threat model that our approach addresses.

\section{Threat Model}
Audio \Captcha{}s are designed to prevent automated adversaries from bypassing access control mechanisms deployed by web services~\citep{szegedy2017inception,he2016deep,hunt2014artificial,fanelle2020blind,arkoselabsArkoseMatchKey}.
We consider an attacker whose goal is to automatically solve audio \Captcha{} challenges at scale to enable automated account creation, credential stuffing, or other abuse.

\smallskip
\noindent \textbf{Attacker Capabilities.}
We assume a black-box setting where the attacker can obtain audio challenges through repeated interactions with the target website, a realistic assumption given that many \Captcha{} services are publicly accessible~\citep{gao2010audio,fanelle2020blind,alnfiai2020novel}. The attacker can leverage off-the-shelf ASR models and advanced LALMs to process and solve the audio challenges. We assume the attacker has knowledge of the general structure of the audio \Captcha{} (e.g., content-based vs.\ rule-based) but does not have access to the internal implementation or training data of the \Captcha{} generation system.

\smallskip
\noindent \textbf{Out of Scope.}
We do not consider adaptive attacks where adversaries fine-tune models specifically on sine-wave speech data, though we discuss this as an important direction for future work in our limitations (\Cref{sec:limitations}). We also do not consider attacks that bypass the \Captcha{} system entirely (e.g., through human solving farms or browser automation vulnerabilities), as these are orthogonal to the audio \Captcha{} design.

To systematically evaluate audio \Captcha{} security under this threat model, we next introduce our evaluation framework and proposed defense.

\section{\IllusionAudio{} Framework}
\label{sec:our_framework}

We introduce \IllusionAudio, a unified framework comprising two components:
(i) \IllusionAudioEval, an evaluation framework for systematically assessing the robustness of audio \Captcha{} schemes against modern LALM-based and ASR-based solvers; and
(ii) \IllusionAudioScheme, a novel audio \Captcha{} design that leverages audio illusion effects.

\subsection{\IllusionAudioEval: Evaluation Framework}
\label{sec:empirical_study}

As shown in \Cref{fig:ACEval}, \IllusionAudioEval employs two AI-driven solvers based on LALMs and ASR models.

\smallskip
\noindent \textbf{LALM-based Solver.} The LALM-based solver employs LALMs to solve audio \Captcha{} challenges in an end-to-end manner.
Given an audio \Captcha{}, the solver feeds the raw audio signal into the LALM, which performs audio perception and semantic reasoning.
For content-based audio \Captcha{}s, the LALM recognizes spoken content, salient acoustic cues, or contextual semantics.
For rule-based audio \Captcha{}s, the model infers and applies the implicit rules specified by the challenge.
The LALM produces a natural-language response describing its interpretation, from which the final answer is extracted automatically. We instantiate the LALM-based solver using three advanced models: Qwen-Audio-Chat, SeaLLMs-Audio-7B, and Qwen2-Audio-7B-Instruct.

\smallskip
\noindent \textbf{ASR-based Solver.} The ASR-based solver adopts a modular pipeline that decouples speech recognition from semantic reasoning. Given an audio \Captcha{}, the solver first transcribes the audio into text using an ASR model. We consider both prompt-guided and non-prompt-guided transcription settings, where prompts optionally provide task-specific hints to improve transcription accuracy.
The resulting transcript is then passed to a downstream LLM (GPT-4o), which performs semantic parsing and reasoning.
For content-based audio \Captcha{}s, the solver extracts the relevant spoken content; for rule-based schemes, it interprets the embedded rules and derives the answer. 
We employ two of the best performing ASR models by the time of writing this paper~\footnote{According to OpenAI~\cite{gpt4otranscribe}, the GPT-4o-Transcript model has lower word error rate and better language recognition accuracy than the original Whisper models. In addition, according to a third party benchmark~\cite{asrcomparison}, GPT-4o-Transcript has the lowest WER and CER among all evaluated models.}: GPT-4o-Transcript and GPT-4o-mini-Transcript.

Using \IllusionAudioEval, we evaluate seven deployed audio \Captcha{} schemes and conduct a user study with 63 participants (\Cref{sec:experiments}). Our findings reveal that existing audio \Captcha{}s are simultaneously vulnerable to AI-driven solvers and difficult for human users. To address these limitations, we propose \IllusionAudioScheme, a new audio \Captcha{} approach based on audio illusions. 
\subsection{\IllusionAudioScheme: Our Method}
\label{subsec: our_method}

\begin{figure*}[t]
\centering
    \begin{subfigure}{0.44\textwidth}
        \centering
        \includegraphics[width=1\textwidth]{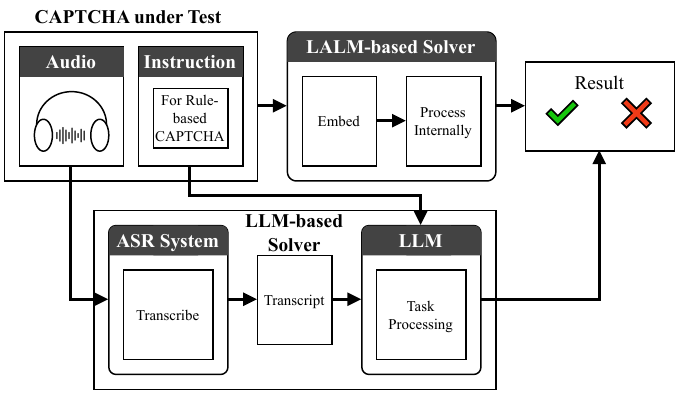}
        \subcaption{\IllusionAudioEval.}
    \label{fig:ACEval}
    \end{subfigure} \hfill
    \begin{subfigure}{0.55\textwidth}
    \centering
    \includegraphics[width=\linewidth]{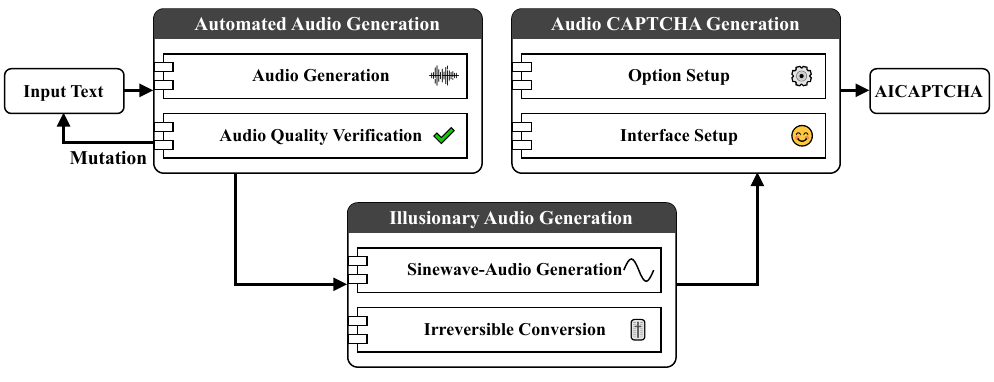}
    \subcaption{\IllusionAudioScheme.}
     \label{fig:illustrationAudio}
    \end{subfigure}
    \caption{Overview of \IllusionAudio.}
    \label{fig:Illusion-based}
\end{figure*}

As illustrated in \Cref{fig:Illusion-based}, \IllusionAudioScheme{} comprises three modules: Automated Audio Generation, Illusionary Audio Generation, and Audio \Captcha{} Generation. We describe each module below.

\smallskip
\noindent
\textbf{Automated Audio Generation.} This module efficiently generates a dataset of short, intelligible audio samples from text prompts without manual annotation.

Starting from an initial prompt \( p_0 \), we iteratively generate candidate audio clips using a text-to-speech (TTS) model~\citep{huggingfaceKokoroHugging}. At iteration \( t \), we synthesize \( K \) candidate clips: \( \mathcal{X}_t = \{ f_{\mathrm{TTS}}(p_t, \theta, \xi_k) \}_{k=1}^{K} \), where \( \xi_k \) are random seeds that ensure diversity and reproducibility. Clips exceeding 2 seconds are discarded. Each remaining clip is assigned an intelligibility score \( s_t(\mathcal{X}_t^k) = s(\mathcal{X}_t^k; p_t) \), which evaluates whether the clip meets human listening requirements, including loudness and tonal quality. Each clip is also transcribed by an ASR model to produce \( \hat{y}(\mathcal{X}_t^k) \). The transcription is compared with the prompt \( p_t \), and pronunciation differences are used to iteratively refine the generation process.

We retain clips satisfying \( s_t(\mathcal{X}_t^k) \geq \tau \), where \( \tau \) is an intelligibility threshold, and add them to the dataset \( \mathcal{D} \). This process repeats until the target dataset size \( |\mathcal{D}| \geq N_{\text{target}} \) is reached. The iterative refinement based on intelligibility scores and transcription feedback ensures high-quality audio samples.


\begin{algorithm}[t]
\caption{Automated Audio Generation}
\label{alg:refined}
\begin{algorithmic}[1]
\Require
initial prompt $p_0$; candidates $K$; target size $N_{\text{target}}$; max duration $T_{\max}{=}2\,\mathrm{s}$;
threshold $\tau$; refinement budget $R$.
\State $\mathcal{D}\gets\varnothing$; \quad $t\gets 0$; \quad $p_t\gets p_0$
\Loop
  \State $\mathcal{X}_t \gets \{\, f_{\mathrm{TTS}}(p_t;\theta,\xi_k)\,\}_{k=1}^{K}$
  \State $\mathcal{X}_t \gets \{\, x \in \mathcal{X}_t \mid T(x)\le T_{\max}\,\}$

  \ForAll{$x \in \mathcal{X}_t$}
    \State $\hat{y}(x) \gets g_{\mathrm{ASR}}(x)$
    \State $s_t(x) \gets s(x; p_t)$
  \EndFor

  \State $\mathcal{X}_t^{\text{good}} \gets \{\, x \in \mathcal{X}_t \mid s_t(x)\ge \tau \,\}$
  \State $m_t \gets N_{\text{target}} - |\mathcal{D}|$
  \State $\mathcal{D} \gets \mathcal{D} \cup \textsc{Take}(\mathcal{X}_t^{\text{good}},\, m_t)$

  \If{$t < R$}
    \State $\Phi_t \gets \textsc{Feedback}\!\left(\{(\hat{y}(x),x)\}_{x\in\mathcal{X}_t^{\text{good}}}\right)$
    \State $p_{t+1} \gets \mathcal{R}(p_t,\Phi_t)$
  \Else
    \State $p_{t+1} \gets p_t$
  \EndIf

  \State $t \gets t+1$; \quad $p_t \gets p_{t+1}$
\EndLoop
\State \Return $\mathcal{D}$
\end{algorithmic}
\end{algorithm}

\smallskip
\noindent
\textbf{Illusionary Audio Generation.}
We employ a sine-wave speech transformation that reduces natural speech to a small set of time-varying sinusoids while preserving the global temporal envelope and coarse formant trajectories. This representation remains intelligible to humans but significantly degrades AI model performance, creating a stable human--AI gap that our audio \Captcha{} exploits.

Formally, given an audio signal \( x \in \mathbb{R}^{L} \), we generate its sine-wave surrogate as:
\begin{equation}\tag{1}
\tilde{x} = \mathcal{S}_{\text{sine}}(x; \psi),
\end{equation}
where \( \mathcal{S}_{\text{sine}} \) denotes the sine-wave renderer with analysis--synthesis parameters \( \psi = (\text{window size}, \text{hop length}, \text{number of formants}) \). The resulting \( \tilde{x} \) serves as our illusionary audio.

Although \( \tilde{x} \) retains human intelligibility, deterministic sine-wave speech could potentially be inverted by specialized reconstruction attacks. To mitigate this risk, we apply a randomized \emph{Irreversible Conversion} module that perturbs the signal through random downsampling:
\begin{equation}\tag{2}
\hat{x} = \mathcal{M}(\tilde{x}; \phi), \qquad \phi \sim \Pi,
\end{equation}
where \( \mathcal{M} \) applies downsampling to \( \tilde{x} \), \( \phi \in [0.5, 0.8] \) is the downsampling factor, and \( \Pi \) is a uniform distribution over this range. The information loss from downsampling prevents reconstruction of the original signal while preserving human intelligibility.
The resulting non-invertible audio signals \( \hat{x} \) form the illusionary audio dataset \( \hat{\mathcal{D}} \) used for \Captcha{} generation.

\smallskip
\noindent
\textbf{Audio \Captcha{} Generation.} Using the clean-audio dataset $\mathcal{D}$ and the illusionary dataset $\hat{\mathcal{D}}$, we adopt an option-based interaction where users select from a small set of audio clips rather than typing a transcript. This design, consistent with industry practice (e.g., Arkose Labs), improves accessibility for users with limited typing proficiency.

A naive multiple-choice formulation introduces three challenges:
(1) \emph{Guessability}: elevated success rates from random selection;
(2) \emph{Low-level heuristics}: solvers may exploit cues such as root mean square (RMS) amplitude; and
(3) \emph{Pattern leakage}: fixed option templates may become learnable.

To address these issues, we randomize both the task framing and option composition for each challenge.
Users identify the illusionary audio that matches the linguistic content of a clean reference.
Candidate options are dynamically sampled, and the unmodified clean audio is sometimes included to break amplitude-based heuristics.
This design increases difficulty for automated solvers while keeping the task straightforward for humans.

Additionally, we support both full-length and segment-wise playback, allowing users to replay only relevant portions rather than entire clips, thereby improving usability.

\section{Evaluation}
\label{sec:experiments}

Our evaluation addresses three research questions:

\begin{itemize}[leftmargin=*, topsep=0pt, itemsep=0pt]
\item \textbf{RQ1: Robustness:} How effectively can \IllusionAudioScheme resist automated attacks from advanced LALM-based and ASR-based solvers?
\item \textbf{RQ2: Usability:} Does \IllusionAudioScheme remain user-friendly and easy to solve for human users?
\item \textbf{RQ3: Ablation Study:} How do some important components of \IllusionAudioScheme contribute to defending against automated attacks?
\end{itemize}

\subsection{Experiment Setup}
\label{subsec:experiment_setup}

\smallskip
\noindent \textbf{Baselines.} We evaluate seven widely deployed audio \Captcha{}s: four content-based schemes (Geetest~\citep{geetestFloat}, Google~\citep{googleReCAPTCHADemo}, MTCaptcha~\citep{mtcaptchaDemoMTCaptcha}, and Telephone-Audio~\citep{githubGitHubLepturecaptcha}) and three rule-based schemes (Math~\citep{fanelle2020blind}, Character~\citep{fanelle2020blind}, and Arkose Labs~\citep{arkoselabsArkoseMatchKey}):

\begin{itemize}[leftmargin=*, topsep=2pt, itemsep=1pt]
\item \textbf{Geetest}: Presents spoken numbers without background noise; users transcribe what they hear.
\item \textbf{Google}: Uses distorted speech with background noise for transcription.
\item \textbf{MTCaptcha}: Adds music as background noise to spoken numbers.
\item \textbf{Telephone-Audio}: Simulates telephone-quality audio with background noise.
\item \textbf{Math}: Embeds a spoken math problem that users must solve.
\item \textbf{Character}: Asks users to count occurrences of a specific digit.
\item \textbf{Arkose Labs}: Requires users to identify and classify audio content.
\end{itemize}

\begin{table*}[t]
\centering
\caption{AI bypass rates of \IllusionAudioScheme{} and existing audio \Captcha{}s against LALM-based and ASR-based solvers. Lower bypass rate indicates stronger security.}
\label{tab:rq1_rq3_merged}
\resizebox{0.96\textwidth}{!}{%
\begin{tabular}{cc|cccccc|cccc}
\hline
\multicolumn{2}{c|}{\textbf{Method}} & \multicolumn{6}{c|}{\textbf{LALM}} & \multicolumn{4}{c}{\textbf{ASR/ASR-LLM}} \\ \hline
\multicolumn{2}{c|}{\textbf{Prompt-Mode}} &
\multicolumn{3}{c|}{\textbf{Zero-Shot}} &
\multicolumn{3}{c|}{\textbf{Chain-of-Thought}} &
\multicolumn{2}{c|}{\textbf{Non-Prompt-Guide}} &
\multicolumn{2}{c}{\textbf{Prompt-Guide}} \\ \hline
\multicolumn{2}{c|}{\textbf{Model-Name}} &
\multicolumn{1}{c|}{\textbf{Qwen}} &
\multicolumn{1}{c|}{\textbf{SeaLLMs}} &
\multicolumn{1}{c|}{\textbf{Qwen2}} &
\multicolumn{1}{c|}{\textbf{Qwen}} &
\multicolumn{1}{c|}{\textbf{SeaLLMs}} &
\textbf{Qwen2} &
\multicolumn{1}{c|}{\textbf{GPT4o-mini}} &
\multicolumn{1}{c|}{\textbf{GPT4o}} &
\multicolumn{1}{c|}{\textbf{GPT4o-mini}} &
\textbf{GPT4o} \\ \hline

\multicolumn{1}{c|}{\multirow{4}{*}{\textbf{\shortstack[c]{Content-based \\ Audio CAPTCHA}}}} &
\textbf{Geetest}   &
93.33\% & 96.66\% & 100.00\% &
96.66\% & 100.00\% & 100.00\% &
100.00\% & 100.00\% &
100.00\% & 100.00\% \\
\multicolumn{1}{c|}{} &
\textbf{Google}    &
76.67\% & 80.00\% & 96.67\% &
83.33\% & 80.00\% & 96.67\% &
70.00\% & 80.00\% &
83.33\% & 80.00\% \\
\multicolumn{1}{c|}{} &
\textbf{MTCaptcha} &
46.66\% & 16.66\% & 30.00\% &
43.33\% & 13.33\% & 33.33\% &
10.00\% & 10.00\% &
16.66\% & 16.66\% \\
\multicolumn{1}{c|}{} &
\textbf{Telephone} &
6.67\% & 3.33\% & 6.67\% &
10.00\% & 3.33\% & 3.33\% &
10.00\% & 10.00\% &
20.00\% & 30.00\% \\ \hline

\multicolumn{1}{c|}{\multirow{3}{*}{\textbf{\shortstack[c]{Rule-based \\ Audio CAPTCHA}}}} &
\textbf{Math}      &
23.30\% & 0.00\% & 0.00\% &
63.33\% & 10.00\% & 3.33\% &
13.33\% & 20.00\% &
23.33\% & 23.33\% \\
\multicolumn{1}{c|}{} &
\textbf{Character} &
16.67\% & 16.67\% & 86.67\% &
26.67\% & 16.67\% & 86.67\% &
93.33\% & 93.33\% &
93.33\% & 96.66\% \\
\multicolumn{1}{c|}{} &
\textbf{Arkoselabs} &
13.33\% & 3.33\% & 10.00\% &
46.66\% & 3.33\% & 6.66\% &
23.33\% & 23.33\% &
30.00\% & 30.00\% \\ \hline

\rowcolor{lightgreen}
\multicolumn{1}{c|}{\textbf{\shortstack[c]{Our Method}}} &
\textbf{\IllusionAudioScheme} &
0.00\% & 0.00\% & 0.00\% &
0.00\% & 0.00\% & 0.00\% &
0.00\% & 0.00\% &
0.00\% & 0.00\% \\
\hline

\end{tabular}%
}
\end{table*}

\begin{table}[t]
\centering
\caption{Human success rates for \IllusionAudioScheme{} and existing audio \Captcha{}s by number of attempts. Higher first-attempt success rate indicates better usability.}
\scalebox{0.66}{
\addtolength{\tabcolsep}{-1ex}
\begin{tabular}{cc|cccc}
\hline
\multicolumn{2}{c|}{\textbf{Method}} & \multicolumn{4}{c}{\textbf{Human Participants}} \\ \hline
\multicolumn{2}{c|}{\textbf{Attempt Times}} 
& \multicolumn{1}{c|}{\textbf{One}} 
& \multicolumn{1}{c|}{\textbf{Two}} 
& \multicolumn{1}{c|}{\textbf{Three}} 
& \textbf{$>$ Three} \\ \hline

\multicolumn{1}{c|}{\multirow{4}{*}{\textbf{\shortstack[c]{Content-based \\ Audio CAPTCHA}}}} 
& \textbf{Geetest}      & 100.00\% & 0.00\%  & 0.00\%  & 0.00\%  \\
\multicolumn{1}{c|}{} 
& \textbf{Google}       & 23.33\% & 33.33\% & 26.67\% & 16.67\% \\
\multicolumn{1}{c|}{} 
& \textbf{MTCaptcha}    & 80.00\%  & 20.00\% & 0.00\%  & 0.00\%  \\
\multicolumn{1}{c|}{} 
& \textbf{Telephone}    & 23.33\% & 36.67\% & 30.00\% & 10.00\% \\ \hline

\multicolumn{1}{c|}{\multirow{3}{*}{\textbf{\shortstack[c]{Rule-based \\ Audio CAPTCHA}}}} 
& \textbf{Math}         & 80.00\%  & 10.00\% & 10.00\% & 0.00\%  \\
\multicolumn{1}{c|}{} 
& \textbf{Character}    & 50.00\%  & 16.67\% & 33.33\% & 0.00\%  \\
\multicolumn{1}{c|}{} 
& \textbf{Arkoselabs}   & 76.67\%  & 6.67\%  & 3.33\%  & 13.33\% \\ \hline

\rowcolor{lightgreen} 
\multicolumn{1}{c|}{\textbf{Our Method}} 
& \textbf{\IllusionAudioScheme}
& 100.00\% & 0.00\% & 0.00\% & 0.00\% \\ \hline

\end{tabular}%
}
\label{tab:human_attempts_combined}
\end{table}

\begin{table*}[t]
\centering

\caption{\IllusionAudioScheme's robustness under different settings.
\IllusionAudioScheme w/o Irreversible-Conversion disables the irreversible conversion module (i.e., keeps the sine-wave transformation but removes the downsampling-based conversion), while \IllusionAudioScheme follows the default design with irreversible conversion enabled.}
\scalebox{0.64}{
\addtolength{\tabcolsep}{-1ex}
\begin{tabular}{cl|cccccc|cccc|c}

\hline
\multicolumn{2}{c|}{\textbf{Method}}
& \multicolumn{6}{c|}{\textbf{LALM}}
& \multicolumn{4}{c|}{\textbf{ASR}}
& \multirow{3}{*}{\textbf{RMS}} \\

\cline{1-12}
\multicolumn{2}{c|}{\textbf{Prompt-Mode}}
& \multicolumn{3}{c|}{\textbf{Zero-Shot}}
& \multicolumn{3}{c|}{\textbf{Chain-of-Thought}}
& \multicolumn{2}{c|}{\textbf{Non-Prompt-Guide}}
& \multicolumn{2}{c|}{\textbf{Prompt-Guide}}
& \\

\cline{1-12}
\multicolumn{2}{c|}{\textbf{Model-Name}}
& \multicolumn{1}{c|}{\textbf{Qwen}}
& \multicolumn{1}{c|}{\textbf{SeaLLMs}}
& \multicolumn{1}{c|}{\textbf{Qwen2}}
& \multicolumn{1}{c|}{\textbf{Qwen}}
& \multicolumn{1}{c|}{\textbf{SeaLLMs}}
& \textbf{Qwen2}
& \multicolumn{1}{c|}{\textbf{GPT4o-mini}}
& \multicolumn{1}{c|}{\textbf{GPT4o}}
& \multicolumn{1}{c|}{\textbf{GPT4o-mini}}
& \textbf{GPT4o}
& \\

\hline
\multicolumn{2}{c|}{\textbf{\IllusionAudioScheme{} w/o Irreversible Conversion}}
& 0.00\% & 0.00\% & \multicolumn{1}{c|}{0.00\%}
& 0.00\% & 0.00\% & 0.00\%
& 0.00\% & \multicolumn{1}{c|}{0.00\%}
& 0.00\% & 0.00\%
& 100.00\% \\ \hline
\rowcolor{lightgreen}
\multicolumn{2}{c|}{\textbf{\IllusionAudioScheme}}
& 0.00\% & 0.00\% & \multicolumn{1}{c|}{0.00\%}
& 0.00\% & 0.00\% & 0.00\%
& 0.00\% & \multicolumn{1}{c|}{0.00\%}
& 0.00\% & 0.00\%
& 0.00\% \\

\hline
\end{tabular}
}
\label{tab:rq3}
\end{table*}

\begin{table}[H]
\centering

    \caption{Human solving attempts for \IllusionAudioScheme under different settings.
\IllusionAudioScheme w/o Clean-Audio removes the clean reference audio, while \IllusionAudioScheme follows the default design with the clean reference played before the illusionary audio.}
\scalebox{0.72}{
\addtolength{\tabcolsep}{-1ex}
\begin{tabular}{ll|cccc}
\hline
\multicolumn{2}{c|}{\textbf{Method}} & \multicolumn{4}{c}{\textbf{Human Participants}} \\ \hline
\multicolumn{2}{c|}{\textbf{Attempt Times}}
& \multicolumn{1}{c|}{One}
& \multicolumn{1}{c|}{Two}
& \multicolumn{1}{c|}{Three}
& >Three \\ \hline
\multicolumn{2}{c|}{\textbf{\IllusionAudioScheme w/o Clean-Audio}}
& 0.00\% & 0.00\% & 3.33\% & 96.67\% \\ \hline
\rowcolor{lightgreen}
\multicolumn{2}{c|}{\textbf{\IllusionAudioScheme}}
& 100.00\% & 0.00\% & 0.00\% & 0.00\% \\ \hline
 \end{tabular}%
}
\label{tab:rq4}
\end{table}

\noindent We generate \Captcha{} samples using official open-source implementations with default configurations. To ensure fair and consistent evaluation, we collect 30 samples per scheme, yielding 210 audio \Captcha{} instances in total.

\smallskip
\noindent \textbf{Metrics.}
We use two complementary metrics. To assess robustness, we report the \emph{bypass rate}, which represents the percentage of \Captcha{} instances successfully solved by AI solvers. A lower bypass rate indicates stronger security. For usability, we report the \emph{success rate}, which represents the percentage of human users who successfully solve the \Captcha{} on each attempt. Higher success rates indicate better accessibility. By comparing AI bypass rates with human success rates, we assess whether a scheme achieves the desired goal of being ``AI-hard'' yet ``human-easy''.

\smallskip
\noindent \textbf{Computation Platform.} All experiments are conducted on a workstation with the Intel Xeon 6 6787P CPU and two NVIDIA RTX 5090 GPUs.

\subsection{Results}

\subsubsection{RQ1: Robustness of Our Method}

We evaluate the robustness of \IllusionAudioScheme{} and seven widely deployed audio \Captcha{} schemes using IllusionAudioEval. The evaluation employs three LALM-based solvers: Qwen-Audio-Chat (Qwen), SeaLLMs-Audio-7B (SeaLLMs), and Qwen2-Audio-7B-Instruct (Qwen2), and two ASR-based solvers: GPT-4o-Transcript (GPT4o) and GPT-4o-mini-Transcript (GPT4o-mini), as described in \Cref{sec:empirical_study}.
For fair comparison, we generate the same number of samples for \IllusionAudioScheme as for the baseline schemes.

\Cref{tab:rq1_rq3_merged} shows that all seven existing schemes exhibit high bypass rates under LALM-based and ASR-based solvers, indicating broad vulnerability to AI-driven attacks.
Content-based \Captcha{}s such as Geetest and Google are easily bypassed by LALMs, with bypass rates near 100\%. ASR-based solvers achieve similarly high bypass rates across most schemes.

In contrast, \IllusionAudioScheme{} achieves a 0\% bypass rate against all tested LALM-based and ASR-based solvers, demonstrating strong robustness against automated attacks.
This superior performance stems from the perceptual asymmetry created by sine-wave speech: the audio remains intelligible to humans but lacks the spectral features that AI models rely on for recognition (\Cref{sec:background}).

\subsubsection{RQ2: Usability of Our Method}

To assess usability, we conduct a user study with 240 audio \Captcha{} instances spanning seven baseline schemes and our method.
We recruit 63 participants, including 36 PVIs and 27 sighted users. Each participant solves 12 distinct \Captcha{}s, yielding 756 total trials. We randomly partition the 240 \Captcha{}s into 20 batches of 12 items, with each batch evaluated by three participants~\footnote{More details about the design of the user study, such as the user interface design of the \Captcha{} instances and the information of the recruited participants can be found in Appendix~\ref{sec:appendix}.}.
For each \Captcha{} instance, we record the maximum number of attempts required by any evaluator as the measure of solving difficulty.

As shown in \Cref{tab:human_attempts_combined}, most existing audio \Captcha{} schemes pose substantial usability challenges.
While Geetest achieves a 100\% first-attempt success rate, other content-based schemes perform worse: Google achieves only 23.33\% and MTCaptcha 80\%.
Similar issues arise for rule-based schemes: Math attains 80\%, but Character and Arkose Labs achieve only 50\% and 76.67\%, respectively.
These results indicate that existing designs impose significant cognitive burdens, often requiring multiple attempts.

In contrast, \IllusionAudioScheme{} achieves a 100\% first-attempt success rate, demonstrating superior usability alongside its strong security properties.

\subsection{RQ3: Ablation Study}

We study the impact of two key design choices in \IllusionAudioScheme{}: the irreversible conversion module and the clean reference audio (\Cref{subsec: our_method}).

\subsubsection{Impact of Irreversible Conversion Module}

We compare \IllusionAudioScheme{} with a variant where the irreversible conversion module is disabled.
To evaluate robustness against lightweight heuristic attacks, we additionally test an RMS-based solver that exploits amplitude patterns.

\Cref{tab:rq3} reports the results. \IllusionAudioScheme{} remains robust against LALM-based and ASR-based attacks regardless of whether irreversible conversion is enabled (0\% bypass rate in both cases).
However, robustness against RMS-based attacks critically depends on this module: without it, the RMS-based attack achieves 100\% bypass rate by exploiting amplitude cues; with irreversible conversion enabled, the attack completely fails (0\% bypass rate).
These results demonstrate that the irreversible conversion module is essential for eliminating low-level amplitude patterns that simple heuristic solvers could exploit.

\subsubsection{Impact of Clean Reference Audio}

We compare our default setting, where clean reference audio is played before the illusionary audio, with a variant that presents only the illusionary audio.

\Cref{tab:rq4} shows a stark contrast. Without the clean reference, 96.67\% of participants required more than three attempts, and none succeeded within the first two attempts.
With the clean reference, all participants (100\%) solved the task on their first attempt.
These results confirm that the clean reference audio is critical for human usability, serving as a perceptual ``priming'' cue that enables listeners to decode the sine-wave speech illusion.

\section{Conclusion}

In this paper, we present \IllusionAudioEval, a novel method to assess audio CAPTCHAs in the AI era. Through extensive evaluation of seven widely deployed audio \Captcha{} schemes, we show that existing designs are vulnerable to AI-driven solvers like Large Audio Language Models and Automatic Speech Recognition systems, which easily bypass them with high success rates. Additionally, we observe significant usability challenges, as many audio CAPTCHAs are not only easily solved by AI but are also difficult for humans, especially those with disabilities. These findings reveal the limitations of current systems, which either fail to thwart AI solvers or impose substantial cognitive burdens on users. In response to these challenges, we propose \IllusionAudioScheme, which exploits the sine-wave speech illusion, a perceptual phenomenon in which speech is encoded into sparse sinusoidal components. While such signals remain intelligible to human listeners, they are much more challenging for AI solvers to interpret due to their lack of conventional acoustic features that AI models typically rely on. By leveraging these perceptual cues, \IllusionAudioScheme creates a significant perceptual gap between humans and machines, ensuring that only humans, regardless of background, can reliably solve the challenge.

\section{Limitations}
\label{sec:limitations}

While our results demonstrate the robustness and usability of \IllusionAudio{} under the evaluated settings, several limitations should be acknowledged.

\smallskip
\noindent \textbf{User Study Scale.}
Although we recruited 63 participants, including both PVIs and sighted users, the sample size limits the statistical power of our analysis. Future studies with larger and more diverse populations would enable more rigorous validation across demographic subgroups.

\smallskip
\noindent \textbf{Controlled Environment.}
Our experiments were conducted in a controlled setting that does not fully capture real-world usage conditions. Factors such as background noise, audio playback devices, hearing ability variations, and network latency may influence both usability and security. Extending the evaluation to more realistic deployment scenarios would provide a more comprehensive assessment.

\smallskip
\noindent \textbf{Adaptive Attacks.}
While \IllusionAudioScheme{} defeats all tested off-the-shelf AI solvers, we did not evaluate against adversaries who specifically fine-tune models on sine-wave speech data. An attacker with access to sine-wave speech examples could potentially train specialized recognition models, leading to an arms race between attack and defense. Possible countermeasures include periodically rotating transformation parameters, combining sine-wave speech with other perceptual phenomena, or incorporating behavioral signals beyond audio recognition. Investigating such adaptive attacks and their mitigations is an important direction for future work.

We view these limitations as natural directions for future research and believe that addressing them will strengthen the practical deployment of perceptually grounded audio \Captcha{} systems.

\bibliography{custom}

@misc{gpt4otranscribe,
	author = {OpenAI},
	title = {Speech-to-text model powered by GPT-4o},
	howpublished = {\url{https://platform.openai.com/docs/models/gpt-4o-transcribe}},
	note = {[Accessed 02-10-2025]},
    year={2025},
}

@misc{asrcomparison,
	author = {Dilmegani, Cem and Alper, Şevval},
	title = {Speech-to-Text Benchmark: Deepgram vs. Whisper in 2026},
	howpublished = {\url{https://research.aimultiple.com/speech-to-text/}},
	note = {[Accessed 02-10-2025]},
    year={2025},
}

@inproceedings{von2003captcha,
  title={CAPTCHA: Using hard AI problems for security},
  author={Von Ahn, Luis and Blum, Manuel and Hopper, Nicholas J and Langford, John},
  booktitle={International conference on the theory and applications of cryptographic techniques},
  pages={294--311},
  year={2003},
  organization={Springer}
}

@inproceedings{ding2025illusioncaptcha,
  title={IllusionCAPTCHA: A CAPTCHA based on visual illusion},
  author={Ding, Ziqi and Deng, Gelei and Liu, Yi and Ding, Junchen and Chen, Jieshan and Sui, Yulei and Li, Yuekang},
  booktitle={Proceedings of the ACM on Web Conference 2025},
  pages={3683--3691},
  year={2025}
}

@inproceedings{gossweiler2009s,
  title={What's up CAPTCHA? A CAPTCHA based on image orientation},
  author={Gossweiler, Rich and Kamvar, Maryam and Baluja, Shumeet},
  booktitle={Proceedings of the 18th international conference on World wide web},
  pages={841--850},
  year={2009}
}

@inproceedings{fanelle2020blind,
  title={Blind and human: Exploring more usable audio $\{$CAPTCHA$\}$ designs},
  author={Fanelle, Valerie and Karimi, Sepideh and Shah, Aditi and Subramanian, Bharath and Das, Sauvik},
  booktitle={Sixteenth Symposium on Usable Privacy and Security (SOUPS 2020)},
  pages={111--125},
  year={2020}
}

@inproceedings{shirali2007captcha,
  title={CAPTCHA for blind people},
  author={Shirali-Shahreza, Mohammad and Shirali-Shahreza, Sajad},
  booktitle={2007 IEEE international symposium on signal processing and information technology},
  pages={995--998},
  year={2007},
  organization={IEEE}
}

@inproceedings{szegedy2017inception,
  title={Inception-v4, inception-resnet and the impact of residual connections on learning},
  author={Szegedy, Christian and Ioffe, Sergey and Vanhoucke, Vincent and Alemi, Alexander},
  booktitle={Proceedings of the AAAI conference on artificial intelligence},
  volume={31},
  number={1},
  year={2017}
}

@inproceedings{he2016deep,
  title={Deep residual learning for image recognition},
  author={He, Kaiming and Zhang, Xiangyu and Ren, Shaoqing and Sun, Jian},
  booktitle={Proceedings of the IEEE conference on computer vision and pattern recognition},
  pages={770--778},
  year={2016}
}

@book{hunt2014artificial,
  title={Artificial intelligence},
  author={Hunt, Earl B},
  year={2014},
  publisher={Academic Press}
}

@article{abdullah2022attacks,
  title={Attacks as defenses: Designing robust audio captchas using attacks on automatic speech recognition systems},
  author={Abdullah, Hadi and Karlekar, Aditya and Prasad, Saurabh and Rahman, Muhammad Sajidur and Blue, Logan and Bauer, Luke A and Bindschaedler, Vincent and Traynor, Patrick},
  journal={arXiv preprint arXiv:2203.05408},
  year={2022}
}

@inproceedings{aubry2025bypassing,
  title={Bypassing Audio reCAPTCHA with Automatic Speech Recognition Models},
  author={Aubry, Paul and Devoivre, Juliette and Carron, Damien and Fernandez, Simon and Duda, Andrzej and Korczy{\'n}ski, Maciej},
  booktitle={2025 IEEE European Symposium on Security and Privacy Workshops (EuroS\&PW)},
  pages={1--5},
  year={2025},
  organization={IEEE}
}

@misc{arkoselabsArkoseMatchKey,
	author = {Arkose Labs},
	title = {Arkose MatchKey (CAPTCHA Software) | Arkose Labs},
	howpublished = {\url{https://www.arkoselabs.com/arkose-matchkey/}},
	year = {2025},
	note = {[Accessed 30-09-2025]},
}

@article{achiam2023gpt,
  title={Gpt-4 technical report},
  author={Achiam, Josh and Adler, Steven and Agarwal, Sandhini and Ahmad, Lama and Akkaya, Ilge and Aleman, Florencia Leoni and Almeida, Diogo and Altenschmidt, Janko and Altman, Sam and Anadkat, Shyamal and others},
  journal={arXiv preprint arXiv:2303.08774},
  year={2023}
}

@article{team2023gemini,
  title={Gemini: a family of highly capable multimodal models},
  author={Team, Gemini and Anil, Rohan and Borgeaud, Sebastian and Alayrac, Jean-Baptiste and Yu, Jiahui and Soricut, Radu and Schalkwyk, Johan and Dai, Andrew M and Hauth, Anja and Millican, Katie and others},
  journal={arXiv preprint arXiv:2312.11805},
  year={2023}
}

@article{Qwen-Audio,
  title={Qwen-Audio: Advancing Universal Audio Understanding via Unified Large-Scale Audio-Language Models},
  author={Chu, Yunfei and Xu, Jin and Zhou, Xiaohuan and Yang, Qian and Zhang, Shiliang and Yan, Zhijie  and Zhou, Chang and Zhou, Jingren},
  journal={arXiv preprint arXiv:2311.07919},
  year={2023}
}

@misc{SeaLLMs-Audio,
    author = {Chaoqun Liu and Mahani Aljunied and Guizhen Chen and Hou Pong Chan and Weiwen Xu and Yu Rong and Wenxuan Zhang},
    title = {SeaLLMs-Audio: Large Audio-Language Models for Southeast Asia},
    year = {2025},
    publisher = {GitHub},
    journal = {GitHub repository},
    howpublished = {\url{https://github.com/DAMO-NLP-SG/SeaLLMs-Audio}},
}

@article{Qwen2-Audio,
  title={Qwen2-Audio Technical Report},
  author={Chu, Yunfei and Xu, Jin and Yang, Qian and Wei, Haojie and Wei, Xipin and Guo,  Zhifang and Leng, Yichong and Lv, Yuanjun and He, Jinzheng and Lin, Junyang and Zhou, Chang and Zhou, Jingren},
  journal={arXiv preprint arXiv:2407.10759},
  year={2024}
}

@misc{mitIllusionsTrick,
	author = {Dana Boebinger and Jarrod Hicks},
	title = {{H}ow do illusions trick the brain? - {M}{I}{T} {M}c{G}overn {I}nstitute},
	howpublished = {\url{https://mcgovern.mit.edu/2022/05/13/use-your-illusion/}},
	year = {2022},
	note = {[Accessed 30-09-2025]},
}

@inproceedings{gao2010audio,
  title={An audio CAPTCHA to distinguish humans from computers},
  author={Gao, Haichang and Liu, Honggang and Yao, Dan and Liu, Xiyang and Aickelin, Uwe},
  booktitle={2010 third international symposium on electronic commerce and security},
  pages={265--269},
  year={2010},
  organization={IEEE}
}

@article{alnfiai2020novel,
  title={A novel design of audio CAPTCHA for visually impaired users},
  author={Alnfiai, Mrim},
  journal={International Journal of Communication Networks and Information Security},
  volume={12},
  number={2},
  pages={168--179},
  year={2020},
  publisher={Kohat University of Science and Technology (KUST)}
}

@article{saini2013review,
  title={A review of bot protection using CAPTCHA for web security},
  author={Saini, Baljit Singh and Bala, Anju},
  journal={IOSR Journal of Computer Engineering},
  volume={8},
  number={6},
  pages={36--42},
  year={2013}
}

@misc{geetestFloat,
	author = {Geetest},
	title = {float --- gt4.geetest.com},
	howpublished = {\url{https://gt4.geetest.com/demov4/voice-float-en.html}},
	year = {2023},
	note = {[Accessed 01-10-2025]},
}

@misc{googleReCAPTCHADemo,
	author = {Google},
	title = {ReCAPTCHA demo --- google.com},
	howpublished = {\url{https://www.google.com/recaptcha/api2/demo}},
	year = {2023},
	note = {[Accessed 01-10-2025]},
}

@misc{mtcaptchaDemoMTCaptcha,
	author = {MTCaptcha},
	title = {Demo | MTCaptcha Code Generator --- service.mtcaptcha.com},
	howpublished = {\url{https://service.mtcaptcha.com/mtcv1/demo/index.html}},
	year = {2023},
	note = {[Accessed 01-10-2025]},
}

@misc{githubGitHubLepturecaptcha, 
  author = {Lepture}, 
  title = {GitHub - lepture/captcha: A captcha library that generates audio and image CAPTCHAs}, 
  howpublished = {\url{https://github.com/lepture/captcha}}, 
  year = {2023}, 
  note = {[Accessed 01-10-2025]}, 
}

@misc{gpt4o-mini,
	author = {OpenAI},
	title = {{O}pen{A}{I} {P}latform — GPT‑4o mini Transcribe Model Documentation},
	howpublished = {\url{https://platform.openai.com/docs/models/gpt-4o-mini-transcribe}},
	year = {2025},
	note = {[Accessed 01‑10‑2025]},
}

@article{gu2024survey,
  title={A survey on llm-as-a-judge},
  author={Gu, Jiawei and Jiang, Xuhui and Shi, Zhichao and Tan, Hexiang and Zhai, Xuehao and Xu, Chengjin and Li, Wei and Shen, Yinghan and Ma, Shengjie and Liu, Honghao and others},
  journal={arXiv preprint arXiv:2411.15594},
  year={2024}
}

@article{ge2023openagi,
  title={Openagi: When llm meets domain experts},
  author={Ge, Yingqiang and Hua, Wenyue and Mei, Kai and Tan, Juntao and Xu, Shuyuan and Li, Zelong and Zhang, Yongfeng and others},
  journal={Advances in Neural Information Processing Systems},
  volume={36},
  pages={5539--5568},
  year={2023}
}

@article{liu2021makes,
  title={What Makes Good In-Context Examples for GPT-$3 $?},
  author={Liu, Jiachang and Shen, Dinghan and Zhang, Yizhe and Dolan, Bill and Carin, Lawrence and Chen, Weizhu},
  journal={arXiv preprint arXiv:2101.06804},
  year={2021}
}

@inproceedings{kulkarni2018audio,
  title={Audio captcha techniques: A review},
  author={Kulkarni, Sushama and Fadewar, Hanumant},
  booktitle={Proceedings of the Second International Conference on Computational Intelligence and Informatics: ICCII 2017},
  pages={359--368},
  year={2018},
  organization={Springer}
}

@inproceedings{sano2013solving,
  title={Solving Google’s continuous audio CAPTCHA with HMM-based automatic speech recognition},
  author={Sano, Shotaro and Otsuka, Takuma and Okuno, Hiroshi G},
  booktitle={International Workshop on Security},
  pages={36--52},
  year={2013},
  organization={Springer}
}

@article{tam2008breaking,
  title={Breaking audio captchas},
  author={Tam, Jennifer and Simsa, Jiri and Hyde, Sean and Ahn, Luis},
  journal={Advances in Neural Information Processing Systems},
  volume={21},
  year={2008}
}

@misc{tiippana2014mcgurk,
  title={What is the McGurk effect?},
  author={Tiippana, Kaisa},
  journal={Frontiers in psychology},
  volume={5},
  pages={725},
  year={2014},
  publisher={Frontiers Media SA}
}

@article{deutsch1974auditory,
  title={An auditory illusion},
  author={Deutsch, Diana},
  journal={Nature},
  volume={251},
  number={5473},
  pages={307--309},
  year={1974},
  publisher={Nature Publishing Group UK London}
}

@incollection{shepherd2017musical,
  title={The musical coding of ideologies},
  author={Shepherd, John},
  booktitle={Whose Music?},
  pages={69--124},
  year={2017},
  publisher={Routledge}
}

@misc{huggingfaceKokoroHugging,
	author = {hexgrad},
	title = {{K}okoro {T}{T}{S} - a {H}ugging {F}ace {S}pace by hexgrad},
	howpublished = {\url{https://huggingface.co/spaces/hexgrad/Kokoro-TTS}},
	year = {2024},
	note = {[Accessed 02-10-2025]},
}

@inproceedings{radzienski2011application,
  title={Application of RMS for damage detection by guided elastic waves},
  author={Radzie{\'n}ski, Maciej and Krawczuk, M and {\.Z}ak, A and Ostachowicz, W and others},
  booktitle={Journal of Physics: Conference Series},
  volume={305},
  number={1},
  pages={012085},
  year={2011},
  organization={IOP Publishing}
}

@misc{googleAICAPTCHA,
	author = {Our Website},
	title = {{A}{I}{C}{A}{P}{T}{C}{H}{A}},
	howpublished = {\url{https://sites.google.com/view/aicaptcha/}},
	note = {[Accessed 02-10-2025]},
    year={2025},
}

@article{wang2025audio,
  title={When Audio and Text Disagree: Revealing Text Bias in Large Audio-Language Models},
  author={Wang, Cheng and Deng, Gelei and Yang, Xianglin and Qiu, Han and Zhang, Tianwei},
  journal={arXiv preprint arXiv:2508.15407},
  year={2025}
}

@article{wang2024audiobench,
  title={Audiobench: A universal benchmark for audio large language models},
  author={Wang, Bin and Zou, Xunlong and Lin, Geyu and Sun, Shuo and Liu, Zhuohan and Zhang, Wenyu and Liu, Zhengyuan and Aw, AiTi and Chen, Nancy F},
  journal={arXiv preprint arXiv:2406.16020},
  year={2024}
}

@article{Li2021RecentAI,
  title={Recent Advances in End-to-End Automatic Speech Recognition},
  author={Jinyu Li},
  journal={ArXiv},
  year={2021},
  volume={abs/2111.01690},
  url={https://api.semanticscholar.org/CorpusID:240419899}
}

@article{yang2024air,
  title={Air-bench: Benchmarking large audio-language models via generative comprehension},
  author={Yang, Qian and Xu, Jin and Liu, Wenrui and Chu, Yunfei and Jiang, Ziyue and Zhou, Xiaohuan and Leng, Yichong and Lv, Yuanjun and Zhao, Zhou and Zhou, Chang and others},
  journal={arXiv preprint arXiv:2402.07729},
  year={2024}
}

@article{Nayeem2025AutomaticSR,
  title={Automatic Speech Recognition in the Modern Era: Architectures, Training, and Evaluation},
  author={Md. Nayeem and Md Shamse Tabrej and Kabbojit Jit Deb and Shaonti Goswami and Md. Azizul Hakim},
  journal={ArXiv},
  year={2025},
  volume={abs/2510.12827},
  url={https://api.semanticscholar.org/CorpusID:282103020}
}

@article{Chorowski2015AttentionBasedMF,
  title={Attention-Based Models for Speech Recognition},
  author={Jan Chorowski and Dzmitry Bahdanau and Dmitriy Serdyuk and Kyunghyun Cho and Yoshua Bengio},
  journal={ArXiv},
  year={2015},
  volume={abs/1506.07503},
  url={https://api.semanticscholar.org/CorpusID:1921173}
}

@article{Bahdanau2014NeuralMT,
  title={Neural Machine Translation by Jointly Learning to Align and Translate},
  author={Dzmitry Bahdanau and Kyunghyun Cho and Yoshua Bengio},
  journal={CoRR},
  year={2014},
  volume={abs/1409.0473},
  url={https://api.semanticscholar.org/CorpusID:11212020}
}

@inproceedings{Ding2025NGCaptchaAC,
  title={NGCaptcha: A CAPTCHA Bridging the Past and the Future},
  author={Ziqi Ding and Shangzhi Xu and Wei Song and Yuekang Li},
  year={2025},
  url={https://api.semanticscholar.org/CorpusID:283934231}
}

@inproceedings{zhang2021generating,
  title={Generating robust audio adversarial examples with temporal dependency},
  author={Zhang, Hongting and Yan, Qiben and Zhou, Pan and Liu, Xiao-Yang},
  booktitle={Proceedings of the Twenty-Ninth International Conference on International Joint Conferences on Artificial Intelligence},
  pages={3167--3173},
  year={2021}
}

@inproceedings{shi2020text,
  title={Text captcha is dead? a large scale deployment and empirical study},
  author={Shi, Chenghui and Ji, Shouling and Liu, Qianjun and Liu, Changchang and Chen, Yuefeng and He, Yuan and Liu, Zhe and Beyah, Raheem and Wang, Ting},
  booktitle={Proceedings of the 2020 ACM SIGSAC conference on computer and communications security},
  pages={1391--1406},
  year={2020}
}

\appendix

\section{Appendix}
\label{sec:appendix}

\begin{figure*}[t]
    \centering
    \begin{minipage}{\textwidth}
        \centering
        \includegraphics[width=0.8\textwidth]{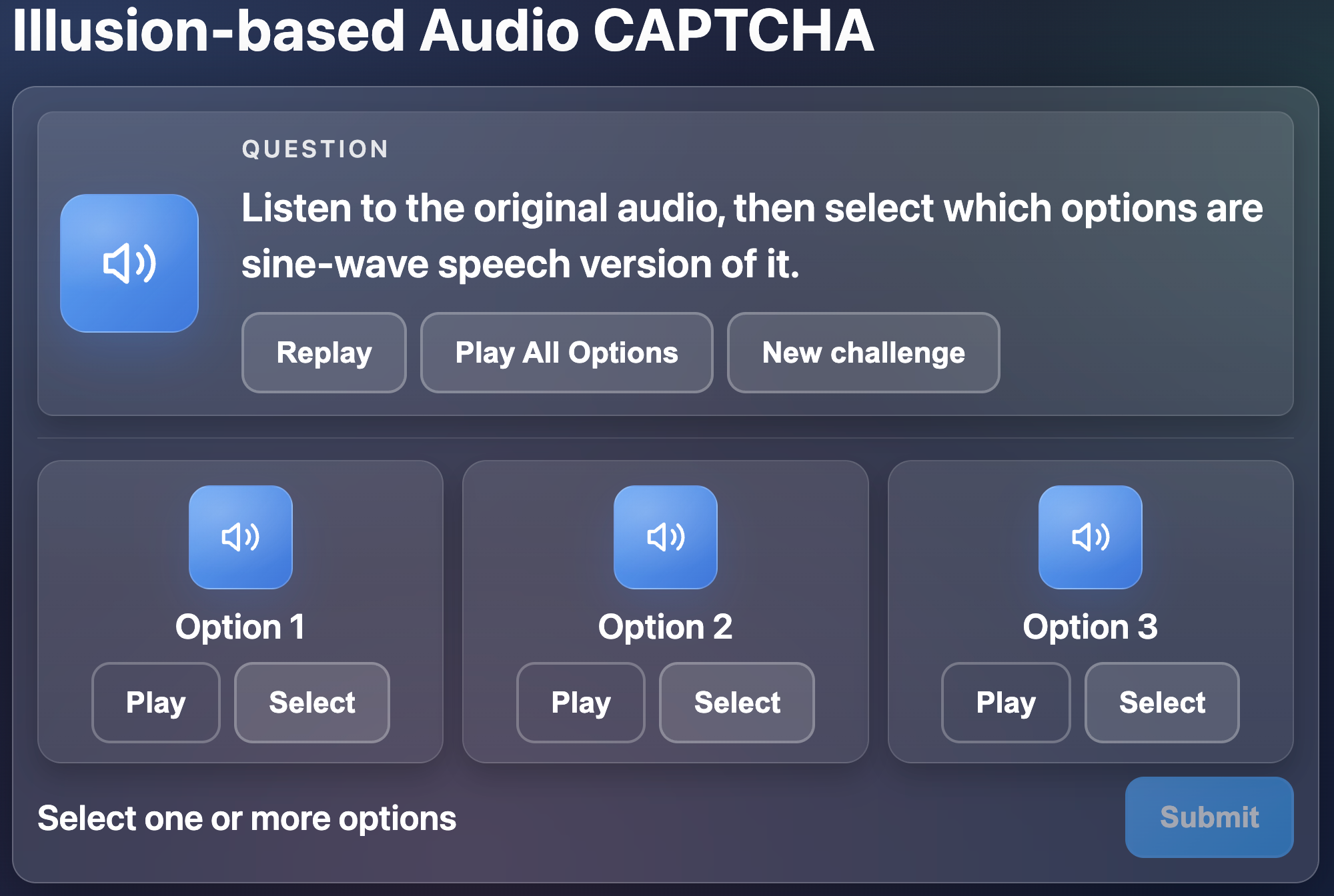}
        \caption{Interface of our \Captcha{}.}
        \label{fig:illusionaudio}
    \end{minipage}
    
    \vspace{1em}
    
    \begin{minipage}{\textwidth}
        \centering
        \includegraphics[width=0.8\textwidth]{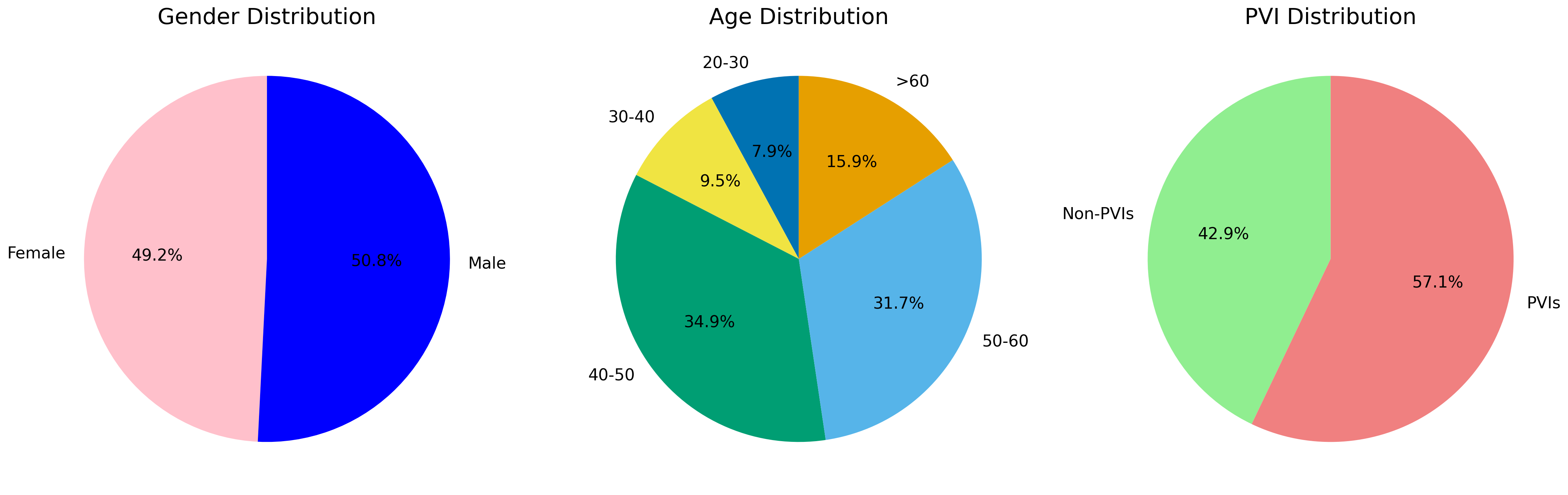}
        \caption{Demographic distributions of participants.}
        \label{fig:demographics}
    \end{minipage}
\end{figure*}

\subsection{CAPTCHA Interface}

The interface of our CAPTCHA system is designed with multiple features to enhance user interaction and accessibility. It supports both random multiple-choice and single-choice options, requiring users to listen to all available choices before making their selection. Additionally, users have the ability to listen to the options sequentially or replay individual options separately. To ensure a smooth experience, a brief training session is provided to introduce basic concepts, such as sine waves. This system advances traditional CAPTCHA designs by engaging deeper levels of human perceptual processing, making it more effective and interactive. Figure \ref{fig:illusionaudio} shows the interface of our CAPTCHA. The design of our CAPTCHA system incorporates key features that enhance user interaction, accessibility, and security. The system supports both random multiple-choice and single-choice options, requiring users to listen to all available choices before selecting an answer. Users can replay or listen to options sequentially, providing greater control over the interaction and reducing frustration. Beyond these interaction mechanisms, the interface is carefully designed to balance usability with security constraints. Visual elements are kept minimal to avoid introducing unnecessary cognitive load, allowing users to focus primarily on auditory perception. Clear and intuitive controls are provided to guide users through the listening and selection process, ensuring that the system remains accessible to users with varying levels of technical proficiency. To further enhance robustness against automated attacks, the system enforces controlled interaction flows. For example, users are required to complete the audio playback before submitting a response, preventing automated agents from skipping directly to an answer. The randomization of option order and playback sequences further increases resistance to pattern-based attacks while preserving a consistent user experience.

\subsection{Participants Information}
Our recruitment process involved thorough documentation of participant information to ensure transparency and ethical use of data. Each participant was required to provide informed consent for their data to be used specifically for the research experiments presented in this paper. As part of the recruitment process, participants were fully informed about the scope of the study and the intended use of their data, and they agreed to participate under these terms. Recruitment was conducted through online video chats, during which we explained the study and monitored participants' behavior throughout the experiment. This approach allowed for real-time interaction, providing valuable insights into their actions and responses.

Throughout both the recruitment and experimental processes, we adhered to all ethical guidelines, prioritizing participant privacy and confidentiality. Personal data was anonymized, and identifying information was kept separate from the experiment data. Participants were informed that they could withdraw from the study at any time without facing any negative consequences, and were assured that their participation was entirely voluntary. Each participant voluntarily chose to take part in the study, with no financial compensation or incentives offered. Their involvement was driven solely by their consent to contribute to the research, ensuring there was no coercion or expectation of rewards for participation.

As shown in Figure~\ref{fig:demographics}, the participant pool exhibits a balanced gender distribution, with male participants accounting for 50.8\% and female participants comprising 49.2\% of the total sample. In terms of age, participants span a wide range, with the majority concentrated between 40 and 60 years old. Specifically, individuals aged 40–50 represent the largest group at 34.9\%, followed by those aged 50–60 at 31.7\%. Younger participants aged 20–30 and 30–40 constitute 7.9\% and 9.5\% of the sample, respectively, while participants over 60 years old account for 15.9\%. Regarding prior exposure to similar systems, 57.1\% of participants fall into the category labeled as PVIs, while the remaining 42.9\% are categorized as Non-PVIs, indicating a diverse range of participant backgrounds relevant to the study. Furthermore, participants are from multiple countries, but most choose not to disclose their country of origin. Therefore, we will not show their origin at this stage. Overall, this demographic composition ensures a heterogeneous participant population, supporting the generalizability of our experimental findings.

\end{document}